\documentstyle[12pt]{article}
\begin{document}
\title{A Note on Gauge Principle and Spontaneous 
Symmetry Breaking in Classical Particle Mechanics}
\author{Naohisa Ogawa 
\thanks{Email: ogawa@particle.sci.hokudai.ac.jp}
\\ Department of Mathematics,  Hokkaido University
\\ Sapporo 060 Japan}
\date{January,  1998}
\maketitle
\begin{abstract}
 The $U(1)$ gauge field is usually induced from the gauge
 principle, that is, the extension of global $U(1)$ phase 
transformation  for matter field. However the phase 
itself is realized only for quantum theory. 
This makes us feel that the gauge fields are induced 
only in quantum theory by gauge principle. To make clear 
this point, we start from the non-relativistic classical 
particle, and we transform its action to the field 
theoretical form classically. We then have  a multiplier 
field which insures the current conservation, and we get a 
global symmetry which translate this multiplier field. 
Using the gauge-principle to extend this global translation 
to the local one, we can introduce the gauge field, and obtain
 the minimal gauge coupling with matter. This shows  that the
 gauge principle is alive for the symmetry of translation 
in classical theory without any phase variable to  obtain 
the gauge field. Using this field theory we show the 
spontaneous symmetry breaking and Higgs mechanism 
which is similar to the super-conductivity. The relation 
of these two phenomenon is discussed.
\end{abstract}
\section{Introduction}
 
  Gauge field is usually explained by gauge-principle for the transformation of phase of matter fields, such as $U(1)$, 
$SU(N)$ gauge fields, and  recently introduced one is the 
higgs field as the gauge field on $Z^2 \otimes M^4$ 
space-time in non-commutative geometry \cite{conne}.
 There are, however, different classes of gauge fields 
which are not related to the quantum mechanical phase.  
Examples are the gravity, spin-connection,  and  recently 
introduced one for classical particle is angular velocity 
which is related to the local (in the time direction) $SO(3)$ transformation \cite{ui}. Another example is the 
geometric gauge field induced by embedding procedure
 \cite{fujii-ohnuki}.
In this paper, we would like to pay attention to the $U(1)$ 
gauge field which are usually introduced by local phase 
transformation in quantum theory, and we seek its origin 
in classical mechanics. To realize this program, first 
order Lagrangian similar to the Polynomial-Formulation 
\cite{fosco} is necessary. The local transformation 
governing gauge-field is the translation of the multiplier 
field which insure the current conservation. The current 
conservation is only the common word to the quantum 
theory for introducing the gauge field but not the phase.
 We introduce $U(1)$ gauge field  by gauge principle in 
section 2, The field equation is solved and we discuss 
its solution in section 3, and   spontaneous symmetry 
breaking and Higgs mechanism is discussed in section 4.  
This field theory is similar to the Schr\"odinger field, 
but is essentially different on the formulation and our 
theory does not come out through the first quantization. 
Our theory does not have an interference effect as 
wave, because it does not include the phase variable 
just having a classical phase-like variable. 
The translation symmetry through which we obtained 
gauge field is spontaneously broken from the outset, 
and we see that the matter field is Nambu-Goldstone 
boson. Higgs mechanism occurs, and the coupled gauge 
field gets mass term which comes from the plasma 
oscillation. 
This is similar to the super conductivity theory.

\section{$U(1)$ gauge field in classical mechanics}

Let us start with  $N$ non-relativistic free particles in 
D-dimensional space by the Lagrangian
\begin{equation}
L({\bf z}(t),\dot{{\bf z}}(t)) = \frac{m}{2} \sum_{k=1}^N 
\dot{\bf z}^2_k.
\end{equation}
Then we can construct the conserved current field as
\begin{equation}
{\bf j}(x,t) = \sum_{k=1}^N \dot{\bf z}_k(t) 
\delta^D(x-z_k(t)),~~~
 \rho(x,t) = \sum_{k=1}^N \delta^D(x-z_k(t)),
\end{equation}
where we take the convention ${\bf z}_k \neq {\bf z}_m$
 if $k \neq m$. 
Using above current field, we can change the free 
particle's action into the form.
\begin{equation}
L = \int d^Dx ~[\frac{m}{2 \rho} ~{\bf j}^2 + 
\alpha(\nabla \cdot {\bf j} + \dot{\rho})],
\end{equation}
where the first term reduces to the original Lagrangian 
by putting the form of current and after the space 
integration. The second term insures the current 
conservation and $\alpha$ is the Lagrange multiplier field. 
For further constraints and hydrodymamical 
interpretation, we discuss in the last section. 
This Lagrangian has a strange form carrying no usual kinetic 
term, but the similar action is well known to treat the 
non-linear sigma model as Polynomial-Formulation 
\cite{fosco}.  Since this action has no local gauge 
invariance, the system is second class and its constraint
 analysis is done straightforward.
The Euler-Lagrange equation leads to three equations
\begin{equation}
\nabla \cdot {\bf j} + \dot{\rho}=0, ~~~\nabla \alpha =
 \frac{m}{\rho} ~{\bf j}, ~~~
\dot{\alpha} = -\frac{m}{2 \rho^2}~ {\bf j}^2.
\end{equation}
By using the second equation of above ones, we can change 
the Lagrangian into the form
\begin{equation}
L = \int d^Dx ~[ -\rho \dot{\alpha} - \frac{\rho}{2 m} 
(\nabla \alpha)^2] .
\end{equation}
This is equal to the Lagrangian for Schr\"odinger field 
\begin{equation}
L = \int d^Dx ~ [i\hbar \Psi^* \partial_t \Psi - \Psi^* 
\hat{H} \Psi],~~~\hat{H} = -\frac{\hbar^2 }{2m} \nabla^2,
\end{equation}
with
\begin{equation}
\Psi \sim \surd \rho(t)~ \exp (\frac{i}{\hbar} \alpha(x,t)) ,
\end{equation}
where we take the approximation that $\rho$ changes 
slowly in space, 
and we have neglected the total derivative. 
From this action we find correspondence between 
Lagrange multiplier field $\alpha$ and  quantum 
mechanical phase variable. 
Now we go back to Lagrangian (3). 
This Lagrangian has the symmetry of global translation 
for multiplier field.
\begin{equation}
 \delta \alpha(x,t) =  C,
\end{equation}
where $C$ is constant. 
The Noether current for this translation is the current 
appearing in action as dynamical variable. Let us extend 
this translation to the local one.
The local translation,
\begin{equation}
 \delta \alpha(x,t) =  \theta(x,t),
\end{equation}
does not change the action only when we extend the 
derivative for
 $\alpha$ field as
\begin{equation}
 \partial_\mu \alpha(x,t)  \to  D_\mu \alpha(x,t) \equiv 
\partial_\mu \alpha(x,t) +  A_\mu (x,t)
\end{equation}
with transformation law for $A_\mu$ is
\begin{equation}
 \delta A_\mu (x,t)  =  - \partial_\mu \theta(x,t).
\end{equation}
We used here the relativistic notation. Notice that this is 
not the usual covariant derivative, and $ D_\mu \alpha(x,t)$
 is not only covariant but also invariant under the local 
translation.
Using the notation
$$ j^\mu = (\rho, j^1, j^2, j^3), $$
our local gauge invariant action takes the form
\begin{equation}
S = \int d^Dx dt~ [\frac{m}{2 \rho} ~{\bf j}^2 - j^\mu A_\mu 
+ \alpha \partial_\mu j^\mu ].
\end{equation}
So we get the usual gauge coupling. Solving the constraint 
(current conservation), we find the solution as the form 
of currents (2), and by putting them into the above action 
we obtain the gauge coupled action for particles.
\begin{equation}
S = \int dt~  \sum_{k=1}^N [\frac{m}{2} \dot{\bf z}_k^2 - 
\dot{z}_k^i \, A_i(z_k,t) - A_0(z_k,t) ].
\end{equation}
In this way we can introduce $U(1)$ gauge field by gauge 
principle completely in classical theory. The local 
symmetry is not the $U(1)$ transformation but 
1-dimensional translation. 
 The reason why the multiplier field corresponds 
to the phase is the following.
In quantum theory the translation of phase induced
the current conservation, and in our case 
the same job is done by the multiplier field.
So they are corresponding each other 
in the sense of gauge transformation.

\section{Field equation}
We consider the free field equation given by the action,
\begin{equation}
S = \int d^Dx dt~[ -\rho \dot{\alpha} - \frac{\rho}{2 m}
 (\nabla \alpha)^2] .
\end{equation}
The Euler-Lagrange equation gives
\begin{eqnarray}
&& \dot{\alpha} + \frac{1}{2m} (\nabla \alpha)^2 = 0,\\
&&\dot{\rho} + \frac{1}{m}\nabla \cdot (\rho \nabla \alpha)=0.
\end{eqnarray}
The solution of the first non-linear equation has the form,
\begin{equation}
\alpha(x,t) = 2m C( \mid \vec{x} - \vec{x_0} \mid - Ct) + D,
\end{equation}
where $\vec{x_0}$, $C$, $D$ are the integration constant.
By using the notation \\
$r=\mid \vec{x} - \vec{x_0} \mid$, the solution of the 
second equation is given as,
\begin{equation}
\rho(x,t) = \int d\omega \frac{A(\omega)}{r^2} 
\exp [i \omega (t-\frac{ r}{2C} )],~~~r \neq 0.
\end{equation}
The green function described by
\begin{equation}
\dot{\rho_G} + \frac{1}{m}\nabla \cdot (\rho_G \nabla 
\alpha)=\delta^3(\vec{x} - \vec{x_0}) \delta(t),
\end{equation}
is given as
\begin{equation}
\rho_G(\vec{x}, \vec{x_0}, t) =
 \frac{(2m)^{-1/2}}{8\pi C r^2} \delta (t-\frac{ r }{2C} ).
\end{equation}
This solution says that particle density moves at the 
constant speed $2C$  which is determined by the initial 
condition. This is essentially classical picture, and much 
different from the Schr\"odinger equation, where the 
equation is diffusion type and propagation belongs to 
that one.
If we work with Schr\"odinger equation, we should add
\begin{equation}
 -\frac{\hbar^2}{8m\rho} (\nabla \rho)^2
\end{equation}
to the action. Then the first equation changes to include 
$\rho$ dependent term to be highly non-linear equation, 
even though the Schr\"odinger equation for $\Psi \sim 
\surd \rho~ \exp (\frac{i}{\hbar} \alpha(x,t)) $ is simple
 linear one.

\section{S.S.B. and Higgs mechanism}
Let us include the gauge field into the action as in section 2.
\begin{equation}
S = \int d^Dx dt~[ -\rho \dot{\alpha} - \frac{\rho}{2 m}
 (\nabla \alpha - {\bf A}(x,t))^2 + \rho A^0] ,
\end{equation}
where we write ${\bf A} = A^k = -A_k$ as usual relativistic 
manner.  The Noether current corresponds to the translation
 invariance is
\begin{equation}
{\bf j} = \frac{\rho}{m}(\nabla \alpha - {\bf A}), ~~~
 j^0 =\rho.
\end{equation}
Therefore the Noether charge $Q$ is given by
\begin{equation}
 Q = \int d^3x ~ \rho (x),
\end{equation}
and induce the infinitesimal translation of $\alpha$.
Let us consider the quantization of this field theory.
The canonical momentum conjugate to density $\rho$ is 
the `` classical-phase" $\alpha$ and we get
\begin{equation}
[\rho(x), \alpha(y)] = i\hbar \delta^3(x-y).
\end{equation}
This relation induce the relation
\begin{equation}
<0\mid [Q, \alpha(x)]\mid 0>=i\hbar = const.
\end{equation}
This means the spontaneous symmetry breaking (SSB), and
 Nambu Goldstone boson is the $\alpha$ field. 
If we add the kinetic term of gauge field, we can perform 
the gauge transformation for gauge field freely.
The gauge transformation
\begin{equation}
A_\mu \to  A'_\mu = A_\mu + \partial_\mu \alpha,
\end{equation}
``gauge out" the $\alpha$ field, and action takes the form 
\begin{equation}
S = \int d^Dx dt~[- \frac{\rho}{2 m} {\bf A}^2(x,t) - \rho A^0 
- \frac{1}{4} F^{\mu\nu} F_{\mu\nu}] .
\end{equation}
$\rho$ is no more dynamical variable. 
The equation for gauge field shows that gauge field gets
 mass term
$$ M = (\frac{\rho \, q^2}{m})^{1/2},$$
where we add the electric charge $q$ to the gauge coupling. 
This is the plasma frequency.  Even if we do not gauge out 
the $\alpha$ field, the electric current has the form
\begin{equation}
{\bf j}_q = -\frac{\bar{\rho} \, q^2}{m}\,({\bf A} - \nabla
 \alpha),
\end{equation}
and taking rotation, we obtain
\begin{equation}
\nabla \times {\bf j}_q = -\frac{\bar{\rho} \, q^2}{m}{\bf B}.
\end{equation}
This is the same as London equation. 
 Here we gave the condition by hand $$\rho = \bar{\rho} + 
\delta \rho \sim \bar{\rho} = const.$$
 So we have the anti-magnetism just like super-conductivity.

\section{Discussion}
Let us discuss on the super-conductivity like solution 
obtained above. Our classical field theory
 should be compared to the charged boson theory rather than 
BCS theory since for fermion $\Psi \sim \surd \rho(t)~ \exp
 (\frac{i}{\hbar} \alpha(x,t))$ can not be applied.  
In the bosonic quantum theory with the Schr\"odinger 
term coupled  with gauge field and screened coulomb repulsion 
term, the repulsive interaction with kinetic term of $\rho$ :(21) determines the coherent length $\xi$ to make the density flat. 
Then we get the London equation in the same way as we have 
explained.  Comparing  to our classical theory, we have no 
 $\nabla \rho$ term (21) and Coulomb repulsion term, the 
effect of them is given by hand as initial condition: 
the flatness of density in our case.  
However we did not pay attention to the stability of London 
equation under other scattering interaction of particle. 
Really we do not have energy gap which insures the stability
 of super-conductivity under other perturbation. \\

    Next we should note another important point which is
 usually discussed on super-conductivity compared to the 
classical mechanics. 
Hereafter we take $q=1$ and $A^0=0$.
In the usual classical theory the electric current satisfies
 the equation under the external electric field.
$$ \frac{\partial \, {\bf j}}{\partial t} \sim \frac{\bar{\rho}}{m}
 {\bf E},$$
or in another form:
$$ \frac{\partial}{\partial t}~[\, {\bf j} + \frac{\bar{\rho}}{m}
{\bf A}] \sim 0.$$
But in London equation (which is the same as our case), total
 time derivative is disappeared. This is the usual discussion
 about difference between classical conductivity and super
 one.
Let us consider carefully on this point.
Using Euler-Lagrange equation of (13) and current form (2)
 explicitly,
we obtain the field theoretical Lorents force.
\begin{equation}
 \frac{\partial \, {\bf j}}{\partial t} = \frac{1}{m}[\, {\rho} {\bf E}
 + {\bf j} \times {\bf B}\,] - \frac{1}{\rho} ({\bf j}\cdot \nabla)
{\bf j},
\end{equation}
where the relation
\begin{equation}
\sum_k \dot{\bf z}_k \, \dot{\bf z}_k \cdot \nabla  \delta^D
(x-z_k) =
\frac{\bf j}{\rho}\,  (\nabla \cdot {\bf j}) = 
\frac{\bf j}{\rho^2} \, ({\bf j}\cdot\nabla \rho) =
\frac{1}{\rho}({\bf j}\cdot\nabla) \, {\bf j}.
\end{equation}
is utilized. The second equality can be interpreted as
\begin{equation}
D_t \rho \equiv \dot{\rho} + {\bf v}\cdot \nabla \rho = 0, 
~~~{\bf v} \equiv \frac{\bf j}{\rho}.
\end{equation}
$D_t$ means the Euler-derivative, and this equation shows 
that the density is preserved along the co-moving frame 
with this fluid. 
This is induced from the current conservation law by using
 non- compressive condition $\nabla \cdot {\bf v} =0$. \\
The third equality can be interpreted as
 $${\bf v} \cdot \nabla {\bf v} = 0.$$
This relation holds in a small class of fluids approximately.\\
  On the other hand if we start from action (12), 
we have Euler-Lagrange equation,
\begin{equation}
{\bf j} = \frac{\rho}{m}(\nabla \alpha - {\bf A}), ~~~ 
\nabla \cdot {\bf j} + \dot{\rho}=0,  ~~~
\dot{\alpha} = -\frac{m}{2 \rho^2}~ {\bf j}^2.
\end{equation}
To obtain the closed equation for current, we need to take 
time derivative of current, and we get
\begin{equation}
 \frac{\partial \, {\bf j}}{\partial t} = \frac{1}{m}[\, {\rho}
 {\bf E} + {\bf j} \times {\bf B}\,] -  {\bf j}\cdot \nabla \,
 (\frac{\bf j}{\rho}) - {\bf j}\,  \frac{\nabla \cdot {\bf j}}{\rho}.
\end{equation}
It seems to be different from the another one, but by using
 the relation (33), we see the equivalence.  We see the last
 two terms in r.h.s. are necessary to form the Euler-derivative
 together with l.h.s. \cite{landau}. 
If we use the variables ${\bf v}$ and $\rho$ for above
 equation, by using the current conservation we obtain
\begin{equation}
m D_t {\bf v} =  {\bf E} + {\bf v} \times {\bf B}.
\end{equation}
This is the well known Navier-Stokes (NS) equation under 
external electromagnetic field without viscosity and 
gradient of pressure.
The conditions (33) are hold for the current expression
 defined by (2),
 but in general it does not hold. Instead starting with the 
action (12) only, we obtain the quite general NS equation. 
Therefore we do not contain (33) as constraints into our 
theory to consider with more general fluids.\\

   Now we come to discuss the main point. 
Even if we start from the action (12), we have the equation of
 motion with Lorents force and only time derivative of current
 are determined as usual. 
But by using the multiplier field $\alpha$, we can write down
 the form of current itself, and obtain the usual (without time
 derivative) London equation. This is a trick by multiplier 
field and this kind of paradox may generally occur in field theory. 
At least for oscillating gauge field, London equation
 holds, and electromagnetic wave gets mass. 
This is well known as the reflection of the ones by ionosphere.
In this phenomena our formulation makes a sense.
\\

\noindent{\em Acknowledgement.}\\
After the first appearance of this article on hep-th, 
the author recieved the comment from Dr.Schakel that he has already 
done the quite similar work  \cite{sch}.
Though his physical view point is different from this article,
 starting Lagrangian is almost the same 
which is originally introduced by C.Eckart \cite{eck}.
He started from the similar effective Lagrangian with internal energy term and  studied the meaning of multiplier field, the relation to non-linear Schroedinger equation, super-fluid and super-conductivity.
 He introduced the gauge coupling to obtain 
 the hydrodynamical-vortex but it is not the electro-magnetic field.
The view point of  ``gauge principle in classical mechanics" is not 
appeared there. This is one of the differences to this article.

The author would like to thank Dr.Nagasawa 
for helpful discussion and encouraging him.
He also thanks Prof.R.Jackiw and Prof.A.M. Schakel for the 
important comments.

\end{document}